\documentclass[reprint,aps, prl, onecolumn,showpacs,groupedaddress,nofootinbib]{revtex4}

\usepackage[utf8]{inputenc}
\usepackage{color}
\usepackage{amsmath,amssymb}
\usepackage{amsfonts}
\usepackage{verbatim}
\usepackage{makeidx}
\usepackage{hyperref}
\usepackage{slashed}
\usepackage[vcentermath]{youngtab}
\usepackage{float}
\usepackage{booktabs}
\usepackage{graphicx}
\usepackage{mathrsfs}
\usepackage{bm}
\usepackage{braket}
\usepackage{url}
\usepackage{etoolbox}
\apptocmd{\sloppy}{\hbadness 10000\relax}{}{}
\allowdisplaybreaks[3]
 \newcommand{\be}{\begin{equation}}
   \newcommand{\ee}{\end{equation}}
     \newcommand{\bea}{\begin{eqnarray}}
   \newcommand{\eea}{\end{eqnarray}}

\usepackage[normalem]{ulem}  
\usepackage{color}
\renewcommand\sout{\bgroup \color{red} \ULdepth=-.5ex \ULset}
\begin{document}

\title{The new resonances ${\bm {Z_{cs}(3985)}}$ and ${\bm {Z_{cs}(4003)}}$  (almost) fill two tetraquark nonets of broken SU(3)$_f$}
\author{Luciano Maiani}
\author{Antonio D. Polosa}
\author{Ver\'onica Riquer}
\affiliation{Dipartimento di Fisica and INFN,  Sapienza  Universit\`a di Roma\\ Piazzale Aldo Moro 2, I-00185 Roma, Italy}
%\pacs{14.40.Rt, 12.39.-x, 12.40.-y}

%\date{\today}
  
\begin{abstract}
New data from BESIII and LHCb indicate the existence of two hidden charm, open strangeness resonances, $Z_{cs} (3985)$ and $Z_{cs}(4003)$. The near degeneracy of $Z_{cs} (3985)$ and $Z_{cs}(4003)$ reproduces, in the strange sector, the situation observed with $X(3872)$ and $Z_c(3900)$.  We  show that, in the tetraquark picture, the $Z_{cs}$ resonances neatly fit into two $SU(3)_f$ nonets with $J^P=1^+$ and opposite charge-conjugation, together with $X(3872)$, $X(4140)$ and $Z_c(3900)$. The mass of the missing element of the nonets is predicted. The classification represents a significative score in favour of the tetraquark model.
\end{abstract} 


\maketitle

Compact tetraquarks are hadrons made by diquark-antidiquark pairs held together by QCD forces in an overall color singlet. As stated in the original proposals and similarly to baryons and $q \bar q$ mesons, tetraquarks form complete multiplets of the flavour symmetry, SU(3)$_f$, broken by the light quark mass differences~\cite{Maiani:2004vq,Maiani:2014aja,Maiani:2016wlq} (for reviews on exotic hadrons see~\cite{Ali:2019roi,Chen:2016qju,Esposito:2016noz,Ali:2017jda,Guo:2017jvc,Lebed:2016hpi,Olsen:2017bmm}. For earlier accounts see also~H.M. Chan~\cite{chan}).
Quark mass differences determine hadron mass differences in two ways: proportionally to the number of valence strange quarks and via the differences in hyperfine, spin-spin interaction, see~\cite{Zeldovich:1967rt,DeRujula:1975qlm,Gasiorowicz:1981jz}.

Flavour SU(3)$_f$ multiplets are not expected for conventional hadronic molecular states, made by color singlet mesons and baryons bound together by forces mediated by color singlet particles (see e.g. the point of view of~\cite{Karliner:2015ina}).
Similarly to what happens in nuclei and due to the large difference in range and strength of pion versus $\eta$ or $\phi$ exchange, prospective molecules not featuring one-pion-exchange forces may not be bound  at all.  
Truly enough, hadron molecular models are still being  considered for hadrons such as $J/\psi$-$\phi $ resonance $X(4140)$ and $Z_{cs}$ where one-pion exchange is obviously impossible~\cite{Aceti:2014uea,Yan:2021tcp,Zhang:2020mpi},  losing contact with nuclear forces. One may  wonder about the physical basis of such models and if they are not simply attempts to simulate the effect of QCD forces in disguise.
The observation of full multiplets of broken SU(3)$_f$ is, in our opinion, a very strong, if not decisive, evidence for compact, QCD based, tetraquark models.

In this note we consider two hidden-charm, strange $J^P=1^+$ resonances recently discovered, namely $Z_{cs}(3985)$, observed  by the BESIII collaboration~\cite{Ablikim:2020hsk,changzheng} in the reaction
\be
e^+e^- \to K^+ Z^-_{cs}(3985)\to  K^+ (D_s^{*-}D^0+D_s^{-}D^{*0})\label{bes}
\ee 
and $Z_{cs}(4003)$ observed by the LHCb Collaboration in $B^+$ decay~\cite{Aaij:2021ivw}
\be
B^+ \to\phi+  Z_{cs}(4003)^+\to \phi + (K^+ J/\psi)\label{lhcb}
\ee
%We show that, together with well established $S$-wave tetraquarks, they fit neatly into two nonets of opposite charge conjugation.

The two particles in \eqref{bes} and \eqref{lhcb} have very different widths and are to be considered as two different states, in spite of being close in mass~\footnote{According to Ref.\cite{Aaij:2021ivw} assuming mass and width of the nominal BESIII result in the amplitude fit to their data "the twice the log-likelihood is worse by 160 units".}. They reproduce in the strange sector the quasi-degeneracy between $X(3872)$  and  $Z_c(3900)$. 

We will show that the two strange resonances can  naturally be classified as the strange components of two $S$-wave tetraquark nonets: (i) the $J^{PC}=1^{++}$ nonet associated with the $X(3872)$ and $X(4140)$; (ii) the $J^{PC}=1^{+-}$ nonet associated with $Z_c(3900)$~\cite{Maiani:2016wlq}.

The $S$-wave tetraquark structures we consider at the start have flavour and spin structures~\cite{Maiani:2004vq,Ali:2019roi} 
 \bea
&&X(3872)=[cq][\bar c\bar q],~X(4140)=[cs][\bar c\bar s]: X, X_{s\bar s}=\frac{1}{\sqrt{2}} \big( | (1, 0)_1 \rangle + | (0, 1)_1 \rangle \big)~(J^{PC}=1^{++})\label{splus} \\
&& Z_c(3900)=[cq][\bar c\bar q]:~\frac{1}{\sqrt{2}} \big( | (1, 0)_1 \rangle - | (0, 1)_1 \rangle \big)~(J^{PC}=1^{+-})  \label{smin}\\
&& Z_c(4020)=[cq][\bar c\bar q]:
\frac{1}{2\sqrt{2}}|(1,1)_1\rangle,~(J^{PC}=1^{+-})\label{sminp}
\eea

$q=u,~d$, diquark and antidiquark spin are indicated and subscripts denote total spin.  We do not consider, for the moment, the nonet associated with $Z_c(4020)$, since $Z_{cs}$ in~\eqref{bes} and \eqref{lhcb} do not qualify to be the strange component of this nonet, due to their mass.

{\emph{\bf {The equal spacing rule.}}
A notable feature of $q\bar q$ meson nonets is that the strange meson mass is half-way between the masses of the non strange and the $s\bar s$ components. The equal spacing rule follows from the flavour symmetry of QCD: flavour  symmetry breaking is due exclusively to quark mass differences. To first order in symmetry breaking, the mass of the hadron is diagonal in the quark basis. For charged mesons, this is guaranteed  by Isospin and Strangeness conservation, while  neutral states separate into the light quark, $I=0$, state and  the  $s\bar s$ state. Thus, hadron masses, up to a constant,  are proportional the mass of the quarks in their composition and are equally spaced according to the number of strange quark  or antiquark. The rule is remarkably satisfied for meson nonets. For the lowest lying vector mesons (we use particle mass values given by Particle Data Group~\cite{pdg}): 
\be
K^*(892)-\frac{\phi(1020)+\rho(775)}{2}\sim 5~{\rm MeV}\label{vectmes}
\ee
The only exceptional case is the pseudoscalar meson nonet, where the mixing of neutral mesons is affected by the ABJ anomaly of the flavour singlet axial current (see~\cite{tHooft:1999cta}). 
Hidden charm, compact tetraquark masses are naturally expected to follow the equal spacing rule, the charm pair acting as spectator flavour singlets.

On the basis of its $J^{PC}=1^{++}$ value, we classify the $J/\psi$-$\phi$ resonance, $X(4140)$, as the $s\bar s$ component of the first nonet ~\cite{Maiani:2016wlq}, a classification supported by the mass difference 
\be
X(4140)-X(3872)=275~{\rm MeV}\sim 2(m_s-m_u)
\ee
To first order in  $SU(3)_f$ symmetry breaking,~the equal spacing rule for the nonet of $X(3872)$ and $X(4140)$ predicts the mass of the strange component
\be
Z_{cs}=\frac{X(4140)+X(3872)}{2}=4009~{\rm MeV} \label{pred1sc}.
\ee 

\emph{\bf {Charge Conjugation for  SU(3)$_f$ nonets.}}
A charge conjugation quantum number can be given to each self conjugate SU(3)$_f$ multiplet according to
\be
{\cal C}T{\cal C}=\eta_T {\tilde T}\label{cconj}
\ee
where ${\cal C}$ denotes the operator of charge conjugation, $T$ the matrix representing the multiplet in $SU(3)_f$ space and ${\tilde T}$ the transpose matrix. $\eta_T$ is the sign taken by neutral members, but it can be attributed, via \eqref{cconj}, to all members of the multiplet.~ %and is conserved in strong and electromagnetic decays, in the exact SU(3)$_f$ limit. 
$\eta=-1$ is given to the electromagnetic current $J^\mu$ while $\eta_{K,\pi}=+1$.  

Eq.~\eqref{cconj} is introduced in the analysis of  ${\cal C}$ conserving meson decays in the limit of exact  $SU(3)_f$, see e.g.~\cite{glashow,kane}. Applied to the trilinear couplings of three meson nonets $A,B,C$ with $\eta_{A,B,C} $ it means that the $SU(3)$ and ${\cal C}$ invariant couplings are obtained from the combinations
\bea
&&D={\rm Tr}\Big(A\{B,C\}\Big),~{\rm for}~\eta_A \eta_B \eta_C=+1\label{DD}\\
&&F={\rm Tr}\Big(A[B,C]\Big),~{\rm for}~\eta_A \eta_B \eta_C=-1\label{FF}
\eea
the two combinations being indicated by $D$ or $F$ couplings in $SU(3)$ jargon~\cite{glashow,Coleman:1985rnk}.

Invariance under transformation~\eqref{cconj} is used to search for restrictions for the e.m. and strong transitions in Eqs.~ \eqref{bes} and \eqref{lhcb} following from flavour $SU(3)$, exact or broken to first order: 
\bea
&& e^+e^-\to T+M,~{\rm (direct~production),~or}: e^+e^-\to Y\to  T+M,~{\rm (resonant~production)}\label{prod}\\ 
&& T\to D^*+\bar D,~T\to M+ J/\Psi,~(T~{\rm decay})\label{decay}
\eea
where  $T$ is one of the tetraquark nonets in~\eqref{splus} and \eqref{smin}, $M $ the pseudoscalar nonet, $Y$ is a resonant state like e.g. $Y(4230)$, that may produce the intermediate step in \eqref{bes}.
Processes (\ref{prod}-\ref{decay}) are described by trilinear couplings and, following Eqs.~(\ref{DD}-\ref{FF}), the charge conjugation of the $T$ nonet selects $D$ or $F$ couplings. 

With the quark electric charge $Q={\rm Diag}(2/3,-1/3,-1/3)$,  we find:
\bea
&&{\cal L}(e^+e^-) \propto{\rm Tr}\Big(Q [X,M]\Big)\propto(-K^+ Z_{cs}^- + K^- Z_{cs}^+),~(J^{PC}=1^{++})\\
&&{\cal L}(e^+e^-) \propto{\rm Tr}\Big(Q \{Z,M\}\Big)\propto \frac{1}{3}(K^+ Z_{cs}^- + K^- Z_{cs}^+),~(J^{PC}=1^{+-})\\
&&{\cal L}(Y_{(I=0,1), s\bar s})\propto (K^+ Z_{cs}^- \mp K^- Z_{cs}^+), {\rm for} ~(J^{PC}=1^{+\pm})
\eea
Direct production in $e^+e^-$ favours the association of $Z_{cs}(3985)$ with the nonet of $X(3872)$ and $X(4140)$. However non strange tetraquarks are more favourably produced by $Y$ resonances and, if this applies to $Z_{cs}$ as well, there would be no preference. A determination of the energy dependence of $Z_{cs}(3985)$ by BES III is crucial, for a comparison with LHCb results. 

%Production and observed decays do not choose to which nonet $Z_{cs}(3985)$ has  to be assigned. 

An indication may come from the decay in~\eqref{decay} observed  by LHCb.  In the exact SU(3)$_f$ limit, with $J/\Psi$ a $SU(3)$ singlet,% the  commutator $[X,M]$ has vanishing trace, while
\bea
&&{\cal M}=\lambda ~\mu ~\psi {\rm Tr}\Big([Z,M]\Big)=0,~ (J^{PC}=1^{++})\label{su3exact}\\
&&{\cal M}=\lambda ~\mu ~\psi {\rm Tr}\{Z,M\}= \lambda ~\mu~\psi (Z_{cs}^+K^-+~c.c.) (J^{PC}=1^{+-})\label{su3exactmin}
\eea
$\lambda$ is a dimensionless coupling, $\mu$ has dimensions of a mass. However, for the $J^{PC}=1^{++}$ nonet, the decay would occur to first order in SU(3)$_f$ symmetry breaking. Denoting by $\epsilon_8 ={\rm Diag}(m_u,~m_d,~m_s)$ the symmetry breaking quark mass matrix
\footnote{ Eqs.~\eqref{su3exact} to \eqref{su3brok} have the same structure as the equations for the mixing of two strange particles in different octets studied in~\cite{kane}. With due account of the charge conjugation of $J/\Psi$, we obtain the same results, for exact or first order broken $SU(3)_f$.}
\be
{\cal M}=\lambda~i \psi {\rm Tr}(\epsilon_8[X,M])\sim \lambda~ (m_s-m_u) ~i \psi( Z_{cs}^+K^- -~ c.c.),~ (J^{PC}=1^{++})\label{su3brok}
\ee
The exact $SU(3)$ limit favours $Z_{sc}(4003)$ in the $Z_c(3900)$ nonet, but the other possibility cannot be excluded.

Finally, the $D ^* \bar D$ decay, \eqref{decay}, is always allowed, and so is the $B^+$ decay  in \eqref{lhcb} (a weak process that requires extending \eqref{cconj} to $CP$ invariance). 
%It is easy to see, however, that CP symmetry does not predict any difference between the two possible classifications of $Z_{cs}$.

In conclusion, lacking information about the energy dependence of \eqref{bes}, we  consider both alternatives.
%@@@@@@@@@@@@@@@@@@@@@@@
 \begin{figure}[htbp]
   \centering
   \includegraphics[width=0.8 \linewidth]{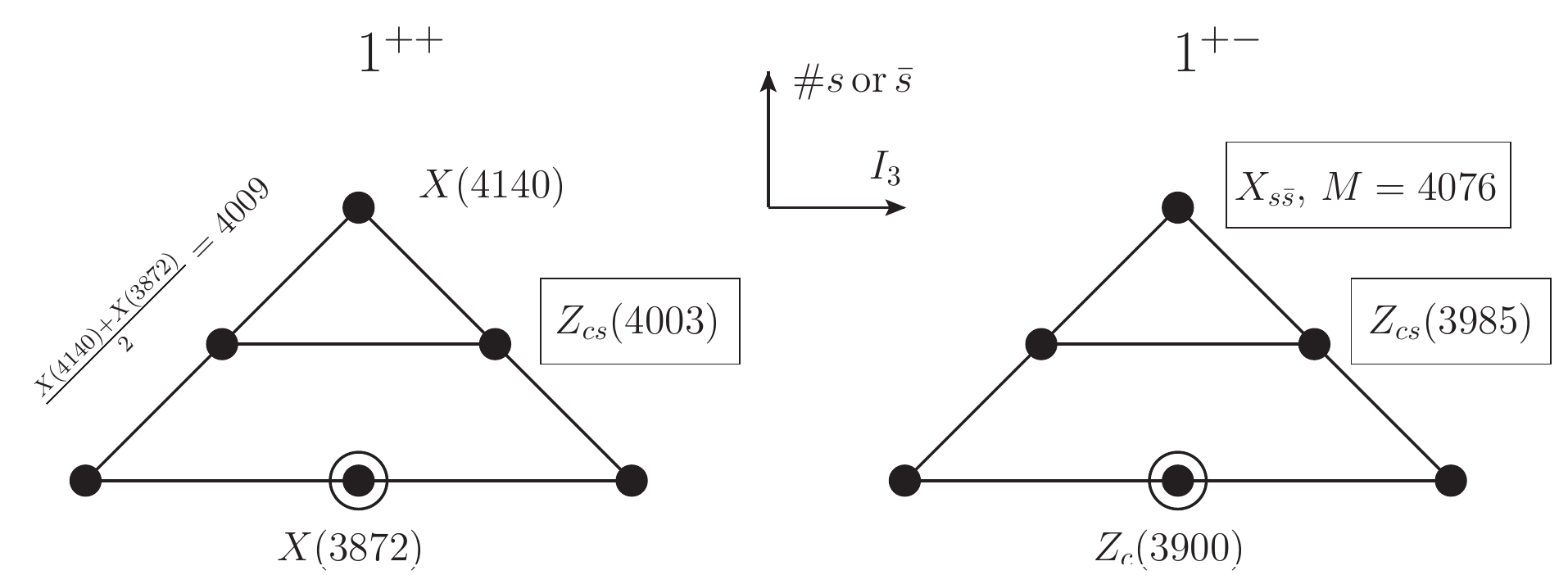}
   \caption{\footnotesize{{\bf Solution 1.}  In the boxes the hidden charm-strange resonances and the missing $X_{s\bar s}$ tetraquark with its predicted mass. The $SU(3)_f$ prediction for the mass of the strange state of the $X(3872) - X(4140)$ nonet is M= 4009~MeV to be compared with the $Z_{cs}$ of Solution1 at 4003~MeV. The  upper state on the right panel has not yet been observed. By ${\cal C}=\pm 1$ nonets we refer to the sign of charge conjugation of the  neutral-non-strange members,~see~\eqref{cconj}. } }

        \label{uno}
\end{figure}
%@@@@@@@@@@@@@@@@@@@@@
t

\emph{\bf {Solution 1.}} The value in~\eqref{pred1sc} supports the assignment of $Z_{cs}(4003)$ to the $J^{PC}=1^{++}$  nonet, see Fig.~\ref{uno}. In this case, the $Z_{cs}$ from BESIII belongs to the  $J^{PC}=1^{+-}$   nonet and the equal spacing rule predicts its  $s\bar s$ component to the mass
\be
X_{s\bar s}(J^P=1^{+-})=4076~{\rm MeV}  \label{pred1ssbar}
\ee
In this solution,  we find a spacing of $275$~MeV in the $C=+1$ nonet, similar to the $\rho-\phi$ spacing ($244$~MeV) and a spacing of $188$~MeV for the $C=-1$ nonet, comparable to the spacing of the tensor mesons $a_2(1320), ~ f^\prime_2(1525)$ ($200$~MeV), still in the range of the strange  to light quark mass difference.

\emph{\bf {Solution 2.}} $Z_{cs}(3985)$ is associated to the $J^{PC}=1^{++}$ nonet, with a disagreement of $\sim 27$~MeV with respect to the equal spacing rule~\eqref{pred1sc}. This is larger than the violation of the rule in the vector meson nonet, \eqref{vectmes}, but  still acceptable. Associating  $Z_{cs}(4003)$ to the  $J^{PC}=1^{+-}$ nonet, we predict
\be
X_{s\bar s}(J^{PC}=1^{+-})=4121~{\rm MeV}  \label{pred2ssbar}.
\ee

In both solutions, one can anticipate that:  
\begin{enumerate} 
\item in case of resonant production in $e^+e^-$, $Z_{cs}(4003)$ should appear also in process~\eqref{bes} as a further peak in the $(D_s^{*-}D^0+D_s^{-}D^{*0})$ spectrum~\cite{future};
\item in the (unlikely) case that production from $e^+e^-$ continuum prevails, production of $Z_{cs}(4003)$ would be suppressed in BES III reaction and Solution 2 is chosen;
\item with improved resolution, LHCb should observe also $Z_{cs}(3085)$;
\item $X_{s\bar s}(J^P=1^{+-})$ should be seen in the decay channels $\eta_c~\phi,~\eta ~ J/\Psi, \bar D_s^*D_s$
\footnote{ A particle heavier than $X(4140)$ with these quantum numbers and decay modes has been anticipated in~\cite{Maiani:2016wlq}.}.
\end{enumerate}

%@@@@@@@@@@@@@@@@@@@@@@@
 \begin{figure}[htbp]
   \centering
   \includegraphics[width=0.8 \linewidth]{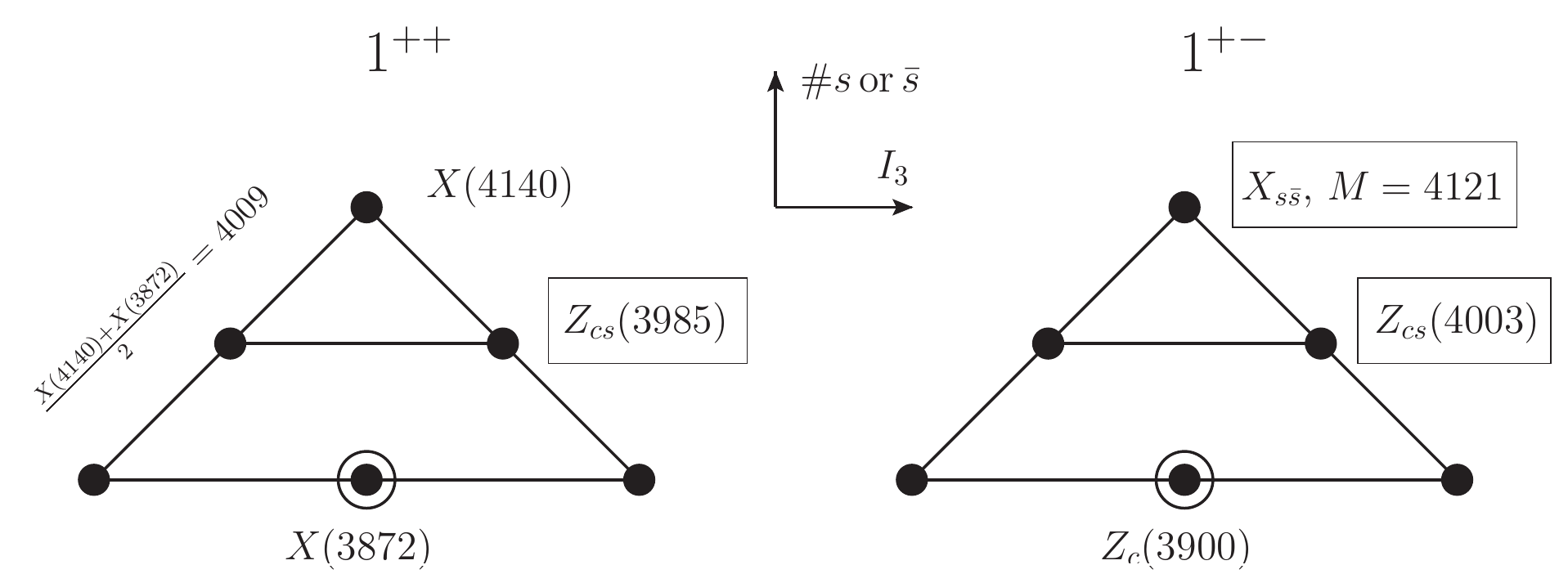}
  \caption{\footnotesize{{\bf Solution 2.}  In the boxes the hidden charm-strange resonances and the missing $X_{s\bar s}$ tetraquark with its predicted mass. The $SU(3)_f$ prediction for the mass of the strange state of the $X(3872) - X(4140)$ nonet is M= 4009~MeV to be compared with the $Z_{cs}$ of Solution1 at 3982~MeV. The  upper state on the right panel has not yet been observed. By ${\cal C}=\pm 1$ nonets we refer to the sign of charge conjugation of the  neutral-non-strange members,~see~\eqref{cconj}.  }}
        \label{due}
\end{figure}
%@@@@@@@@@@@@@@@@@@@@@

The discovery of two hidden charm-strange resonances, almost degerate and in the right mass ballpark represents a %remarkable progress towards the 
confirmation of the tetraquark scheme. Solution 1 is clearly more favourable, but Solution 2 cannot be excluded at the moment. 

Besides confirming the $X_{s\bar s}(J^{PC}=1^{+-})$, much remains to be done.

A third $SU(3)_f$ nonet is expected, associated to $Z_c(4020)$. Therefore, a third $Z_{cs}$ with $J^P=1^+$ is expected, at an estimated mass of $4170-4200$~MeV, and a second $X_{s\bar s}$. LHCb reports observation of a $Z_{cs}(4220)$, with possible $J^P=1^+$ or $1^-$~\cite{Aaij:2021ivw}. We wait for a clarification of spin-parity.

The shopping list towards complete nonets contains, in addition, the  $I=1$ partners of $X(3872)$, decaying into $J/\psi + \rho^\pm$ and the $I=0$ partners of $Z_c(3900)$ and $Z_c(4020)$, possibly decaying into $J/\psi +f_0(500)$(aka $\sigma$).

\section*{Acknowledgements}
%\begin{acknowledgments}
We are grateful to Chang-Zheng Yuan for useful correspondence about exotic hadron production in $e^+ e^-$ annihilation.
%\end{acknowledgments}

\end{document}